\documentstyle[12pt,bezier]{article}
\begin{document}

\title{
Scalar Curvature Factor in the Schr\"odinger Equation and Scattering
on a Curved Surface}  
\author{
Ali Mostafazadeh\thanks{Supported by Killam Postdoctoral Fellowship;
E-mail: alimos@phys.ualberta.ca}\\ \\
Department of Physics, 
Sharif University of Technology\\ 
P.~O.~Box 11365-9161, Tehran, Iran, \\
Institute for Studies in Theoretical Physics and Mathematics\\
P.~O.~Box 19395-1795, Tehran, Iran, and\\
Theoretical Physics Institute, 412 Physics Lab, University of Alberta,\\
Edmonton, AB, Canada T6G 2J1\thanks{Current address.}.
}
\date{ }
\maketitle
\begin{abstract}
The scattering of free particles constrained to move on a cylindrically
symmetric curved surface is studied. The nontrivial geometry of the space
contributes to the scattering cross section through the kinetic as
well as a possible scalar curvature term in the quantum Hamiltonian. 
The coefficient of the latter term is known to be related to the factor 
ordering problem in curved space quantization. Hence, in principle,
the scattering data may be used to provide an experimental resolution 
of the theoretical factor ordering ambiguity. To demonstrate the sensitivity
required of such an experimental setup, the effect of a localized
magnetic field in the scattering process is also analyzed.
\end{abstract}

\newpage

\baselineskip=24pt
\section{Introduction}
One of the oldest and in some cases most severe problems of quantization
of classical systems, is the factor ordering ambiguity. The ordinary
operator quantization of non-relativistic classical systems involves promotion of the coordinate $(x^i)$ and momentum $(p_i)$ variables to
linear operators $\hat x^i$ and $\hat p_i$ acting on a Hilbert space ${\cal H}$ and satisfying the Weyl-Heisenberg algebra. The dynamics is then 
dictated by the Schr\"odinger equation which involves the Hamiltonian operator $\hat H$. The quantum Hamiltonian $\hat H$ is constructed from the
classical Hamiltonian $H$ according to the following requirements:
        \begin{itemize}
        \item[1)] $\hat H$ is a self-adjoint linear operator acting on 
        $\cal H$;
        \item[2)] In the classical limit $\hbar\to 0$, $\hat H\to H$.
        \end{itemize}

In general, these two conditions do not determine $\hat H$ uniquely. Thus
a classical system may lead to different quantum systems. For instance,
consider the motion of a free particle of mass $m$ moving on a Riemannian 
manifold $M$. The classical Hamiltonian is given by:
        \begin{equation}
        H=\frac{1}{2m}g^{ij}(x)\, p_i\, p_j\;,
        \label{q1}
        \end{equation}
where $g^{ij}$ are components of the inverse of the metric tensor:
        \begin{equation}
        \mbox{\large $g$}=g_{ij}(x)\, dx^i\otimes dx^j\;.
        \label{q2}
        \end{equation}
Since $g^{ij}$ depend explicitly on the coordinates $x:=(x^1,\cdots,x^n)$,
the requirements 1) and 2) do not determine $\hat H$ uniquely. In this
case even the additional requirement of form invariance under coordinate
transformations does not lead to a unique choice for $\hat H$ \cite{marinov}.

To quantize a classical Hamiltonian, therefore, one may need to make a choice
of factor ordering. One way of achieving this is to appeal to alternative
quantization schemes which make the choice of factor ordering implicitly.
The best known example of this is the path integral quantization schemes.

In view of the form invariance requirement, different choices of factor ordering for
the system of Eq.~(\ref{q1})  differ by a multiple of $\hbar^2 R$,  where $R$
denotes the scalar
Ricci curvature of $M$. That is, in general, one has:
        \begin{equation}
        \hat H=\frac{1}{2m}\, \hat g^{-1/4}\, \hat p_i\,\hat g^{ij}
        \,\hat g^{1/2}\,\hat p_j\,\hat g^{-1/4}+
        \frac{\lambda\hbar^2}{m}\,\hat{R}\;,
        \label{q0}
        \end{equation}
where $g:=det(g_{ij})$ and $\lambda$ is a real parameter reflecting the
factor ordering ambiguity in quantizing (\ref{q1}). One can express (\ref{q0})
in a local coordinate representation:
        \begin{equation}
        \langle x|\hat H=\left(\frac{-\hbar^2}{2m}\Delta+\frac{\lambda\hbar^2}{m}
        R(x)\right)\langle x|\;.
        \label{q3}
        \end{equation}
Here $\Delta:=g^{ij}\nabla_i\nabla_j$ denotes the Laplace-Beltrami operator
and $\nabla_i$ stands for the covariant derivative corresponding to the
Levi Civita connection along $\partial/\partial x^i$. Choosing the Hilbert
space to be $L^2(M)$, i.e., the state vectors $|\psi\rangle$ to be scalar second 
integrable functions on $M$, one has:
        \begin{equation}
        \langle x|\hat H|\psi\rangle=\frac{\hbar^2}{2m}\left(
        -\frac{1}{\sqrt{g}}\partial_i g^{ij}\sqrt{g}\partial_j+
        2\lambda R\right) \psi(x)\;.
        \label{q4}
        \end{equation}
Note that in general the momentum operators are represented in the
coordinate representation according to:
        \[ \langle x|\hat p_i=\hbar(-i\frac{\partial}{\partial x^i}+
        \omega_i)\langle x|\;,\]
where $\omega_i$ are components of a closed one-form $\omega$ on $M$. For 
convenience we assume $M$ to be simply connected, so that $\omega$ is exact.
In this case it may be gauged away. 

In the long history of the problem of curved space quantization 
\cite{marinov}, there have been different arguments offered in support of
the choices: $\lambda=0$ \cite{cd,k}, $\lambda=1/12$ \cite{bd1,cheng}
(see also \cite{marinov}),
$\lambda=1/8$ \cite{bd2}. To the author's best knowledge none of these
arguments are based on solid factual grounds for the case of general
Riemannian manifolds. 

For Lie group manifolds and homogeneous spaces, there exist 
group theoretical
quantization methods (See \cite{marinov} 
for references). 
But for these cases, $R$ is constant and 
the ambiguity in $\lambda$ 
is physically irrelevant. 
The situation is probably best described by Marinov \cite{marinov} 
who says: ``I wonder, whether it is possible to decide at present between 
the two variants'' [$\lambda=0,1/12$] ``of quantizing the Hamiltonian as long as no nontrivial (i.e., $R\neq$ const.) solvable examples are known.''

The same problem may be addressed for the supersymmetric extensions
of (\ref{q1}). Ref.~\cite{p5} provides a thorough analysis of a 
supersymmetric quantum mechanical system  based on an arbitrary spin manifold.
In this case, the quantum Hamiltonian is
given as the square of the generator of the supersymmetry \cite{p4}. The
quantization 
of the latter is free of factor ordering ambiguity. Therefore the Hamiltonian
operator is unambiguous. Indeed it is given by Eq.~(\ref{q0}) with the
choice $\lambda=1/8$. Furthermore, this choice for $\lambda$ is 
shown to be consistent with the path integral quantization scheme \cite{p5}.
Note however that one may not conclude from the knowledge of a supersymmetric
extension of (\ref{q1}) that $\lambda=1/8$ for the original purely bosonic
theory. Although the form of the Hamiltonian is analogous, the Hilbert
spaces are different. For example, for the system considered in \cite{p5}
and \cite{p4} the Hilbert space is the space of spinors on $M$ and
the $\hat p_i$'s involve components of the spin connection.

The purpose of the present article is to seek physical consequences of 
possible existence of the scalar curvature factor in the Hamiltonian. 
The only known physical effect that causes curvature in three 
dimensional  space is gravity. Thus one might be tempted to address
the question by studying the effect of gravity on non-relativistic 
quantum systems.  It is clear that for the experimentally available 
quantum systems, the effect of the scalar curvature factor due to gravity
must be extremely weak. This is because in addition to being a 2--loop
($\hbar^2$) order effect,  the scalar curvature factor is proportional to
the Ricci scalar curvature. For such systems the latter is caused by 
gravity  and therefore it is extremely small.
An alternative approach to the problem is to study effectively 
two--dimensional systems which are constrained to have dynamics
in a curved surface.  The curvature may in principle be maintained 
mechanically and thus made considerably large.  
In particular, we shall study a simple scattering problem in two dimensions.
The idea is to pave the way for a potential experimental resolution of
this sort of factor ordering ambiguity. 

Consider a free particle moving on a two-dimensional surface $M$ embedded
in ${\relax{\rm I\kern-.18em R}}^3$. For simplicity, suppose that $M$ is asymptotically flat and 
has a cylindrically symmetric geometry and trivial (${\relax{\rm I\kern-.18em R}}^2$) topology. Then
the scattering problem may be formulated as in the case of potential
scattering. The scattering data, however, reflect the nontrivial
geometry of $M$. In particular this involves the contribution of the
scalar curvature factor.

In Sec.~2, we review the basic formalism used in the study of the
scattering problem in two dimensions. Sec.~3 exhibits the treatment of
the specific problem of scattering due to geometry. Sec.~4 treats the
effect of a localized magnetic field in the scattering of
charged particles moving on the curved surface of interest. The result
is used to provide an order of magnitude estimate for the sensitivity
required for observing the contribution of the scalar curvature factor
to the scattering cross section. Sec.~5 includes the concluding remarks.

\section{Scattering in Two Dimensions}
Consider the ordinary time independent potential scattering problem
in two dimensions. The Lippmann-Schwinger equation for the Hamiltonian
        \begin{equation}
        \hat H=\hat H_0+\hat V\;,
        \label{q2.1}
        \end{equation}
is given by:
        \begin{equation}
        |\psi^{(\pm)}\rangle=|\phi\rangle+
        \frac{1}{E-\hat H_0\pm i\epsilon}
        \hat V|\psi^{(\pm)}\rangle\;,
        \label{q2.2}
        \end{equation}
where we follow the notation of Sakurai \cite{sakurai}. In Eqs.~(\ref{q2.1})
and (\ref{q2.2}),
        \begin{equation}
        \hat H_0=\frac{1}{2m}\,\delta^{ij}\,\hat p_i\,\hat p_j 
        \label{q2.0}
        \end{equation}
is the free Hamiltonian and $\hat V$ is the interaction potential.

In the coordinate representation the second term in (\ref{q2.2}) can be 
written in the form:
        \begin{equation}
        \langle x|\frac{1}{E-\hat H_0\pm i\epsilon}\hat V|\psi^{(\pm)}\rangle
        =\frac{m}{2\pi^2\hbar^2}\int d^2x^{'}I^{(\pm)}(x,x',k)\langle x'|
        \hat V|\psi^{(\pm)}\rangle\;,
        \label{q2.4'}
        \end{equation}
where
        \begin{equation}
        I^{(\pm)}:=\int d^2q\,\frac{e^{i\vec q.(\vec x-\vec x')}}{k^2-q^2
        \pm i\epsilon}\;, ~~~~~~ k:=|\vec k|=\frac{\sqrt{2mE}}{\hbar}\,.
        \label{q2.4}
        \end{equation}
As in the three dimensional case, one may switch to the polar coordinates
to evaluate the integrals (\ref{q2.4}). Performing the angular integration
and consulting \cite{gr}, one has:
        \[
        I^{(\pm)}=2\pi\int_0^\infty dq \,
        \frac{q\, J_0(q|\vec x-\vec x'|)}{k^2-q^2\pm i\epsilon}=
        -i\pi^2\, H_0^{(1)}(\pm k|\vec x-\vec x'|+i\tilde\epsilon)\;,
        \]
where $J_0$ and $H_0^{(1)}$ are Bessel and Hankel functions and $\tilde
\epsilon$ is another infinitesimal positive parameter.

For the scattering problem, one considers the large $r:=|\vec x|$
limit. Using the asymptotic properties of the Hankel function \cite{gr},
for $r\gg |\vec x'|$ one finds:
        \[
        \langle x|\frac{1}{E-\hat H_0\pm i\epsilon}\hat V|\psi^{(+)}\rangle=
        \left(
        \frac{-ime^{-i\pi/4}}{\sqrt{2\pi}\hbar^2}\right)\frac{e^{ikr}}{
        \sqrt{kr}}\int d^2x'\,e^{-i\vec k'.\vec x'}\langle x'|\hat V|
        \psi^{(+)}\rangle\;,
        \]
where $\vec k':=k\hat{x}$. The iterative solution of (\ref{q2.1}),
is then carried out by setting $\langle x|\phi\rangle=\langle x|k\rangle
=e^{i\vec k.\vec x}/2\pi$. One then obtains:
        \begin{equation}
        \langle x|\psi^{(+)}\rangle=\frac{1}{2\pi}\left( e^{i\vec k.\vec x}+
        \frac{e^{ikr}}{\sqrt{r}}f(\vec k',\vec k)\right)\;.
        \label{q2.5}
        \end{equation}
In the first Born approximation:
        \begin{equation}
        f(\vec k',\vec k)\approx f^{(1)}(\vec k',\vec k)=
        \frac{-i\sqrt{2\pi}me^{-i\pi/4}}{\sqrt{k}\hbar^2}
        \int dx^{'2}e^{-i\vec k'.\vec x'}\langle x'|\hat V|k\rangle\;,
        \label{q2.6}
        \end{equation}
where:
        \begin{equation}
        \langle x'|\hat V|k\rangle=\frac{1}{2\pi}\int d^2x'' e^{i\vec k.\vec x''}
        \langle x'|\hat V|x''\rangle\;.
        \label{q*}
        \end{equation}
The scattering cross section is related to its amplitude $f$ by the relation:
        \begin{equation}
        \frac{d\sigma(\vec k,\vec k')}{d\Omega}=|f(\vec k,\vec k')|^2
        \;.
        \label{q2.7}
        \end{equation}

In the ordinary potential scattering $\hat V$ is a local operator:
        \[\langle x'|\hat V|x''\rangle=V(x')\delta(\vec x'-\vec x'')\;.\]

Consider expressing the Hamiltonian (\ref{q4}) in the form (\ref{q2.1}).
In view of Eq.~(\ref{q2.0}), this leads to:
\begin{equation} 
\langle x'|\hat V|x''\rangle = 
\frac{\hbar^2}{2m}
\left[
(\delta^{ij}-g^{ij})
\partial'_i\partial'_j-
\frac{\partial'_i(\sqrt{g}g^{ij})}{\sqrt{g}}\partial'_j+
2\lambda R
\right]
\delta(\vec x'-\vec x'')\;,
\label{q2.8}
\end{equation}
where $\partial_i'$ means partial derivation with respect to $x^{'i}$.
The curved space scattering is seen to correspond to an ultralocal 
potential. 

We conclude this section by noting that the correspondence with the potential 
scattering on ${\relax{\rm I\kern-.18em R}}^2$ is justified, for $M$ is assumed to be asymptotically flat
and topologically trivial. The latter allows one to use a single coordinate 
patch in performing the computations.

\section{Scattering Due to a cylindrically Symmetric Geometry}
Consider a surface $M\subset{\relax{\rm I\kern-.18em R}}^3$ defined by the equation:
        \begin{equation}
        z=f(r)\;,
        \label{q3.1}
        \end{equation}
where $(r,\theta,z)$ are cylinderical coordinates on ${\relax{\rm I\kern-.18em R}}^3$ and 
$f:[0,\infty)\to{\relax{\rm I\kern-.18em R}}$ is a smooth function with vanishing first
derivative at the origin, i.e., $\dot f(0)= 0$. This is the
condition which makes $M$ a differentiable manifold. Furthermore,
assume that $f$ has a (physically) compact support. Thus $M$ is 
asymptotically flat.

The implicit geometry of $M$ is described by the (induced) metric
(from ${\relax{\rm I\kern-.18em R}}^3$):
        \begin{equation}
        (g_{ij})=\left(
        \begin{array}{cc}
        F^2&0\\
        0&r^2
        \end{array}\right)\;,
        \label{q3.2}
        \end{equation}
where $F^2:=1+\dot f^2$. Given the metric, one can easily compute
the terms in Eq.~(\ref{q2.8}). In view of Eq.~(\ref{q*}), one then has:
        \begin{eqnarray}
        \langle x'|\hat V|k\rangle&=&\frac{\hbar^2}{4\pi m}\left\{
        (1-\frac{1}{F^2})\frac{\partial^2}{\partial r^{'2}}+
        [\frac{1}{r'}(1-\frac{1}{F^2})-\frac{\dot F}{F^3}]\frac{\partial}{
        \partial r'}+\frac{4\lambda\dot F}{r'F^3}\right\}e^{i\vec k.\vec x'}
        \;,\nonumber\\
        &=&\frac{\hbar^2}{4\pi m}\left\{
        -(1-\frac{1}{F^2})\left(\frac{\vec k.\vec x'}{r^{'2}}\right)^2+
        i[\frac{1}{r'}(1-\frac{1}{F^2})-\frac{\dot F}{F^3}]\left(
        \frac{\vec k.\vec x'}{r'}\right)+\frac{4\lambda\dot F}{r'F^3}
        \right\}e^{i\vec k.\vec x'}\;.\nonumber
        \end{eqnarray}
The latter formula together with Eq.~(\ref{q2.6}) lead to the expression
for $f^{(1)}(\vec k',\vec k)$:
        \begin{eqnarray}
        f^{(1)}(\vec k',\vec k)&=&\frac{e^{-3\pi i/4}}{\sqrt{8\pi k}}
        \int d^2x' e^{i(\vec k-\vec k').\vec x'} \left\{
        -(1-\frac{1}{F^2})
        (\frac{\vec k.\vec x'}{r'})^2+\right.\nonumber\\
        &&\left.
        i[\frac{1}{r'}(1-\frac{1}{F^2})-\frac{\dot F}{F^3}]
        (\frac{\vec k.\vec x'}{r'})+\frac{4\lambda\dot F}{r'F^3}
        \right\} \;
        \label{q90}
        \end{eqnarray}
To perform the integral on the r.\ h.\ s.\ of (\ref{q90}), we choose a
coordinate system in which $\Delta\vec k:=\vec k-\vec k'$ is along
the $x'-$axis. Then switching to polar coordinates $(r',\theta')$, one 
can evaluate the angular integration. This results in:
        \begin{eqnarray}
        f^{(1)}(\vec k',\vec k)&=&\sqrt{\frac{\pi}{2k}}e^{-3\pi i/4}
        \int_0^\infty dr' \left\{
        \left[ 
        -r'(1-\frac{1}{F^2})k_x^2+\frac{4\lambda\dot F}{F^3}
        \right]
        J_0(r'|\Delta\vec k|)+ \right.\nonumber\\
        &&\left.\left[
        \frac{k_y^2-k_x^2}{|\Delta\vec k|}-k_x(1-\frac{1}{F^2}-
        \frac{r'\dot F}{F^3})\right] J_1(r'|\Delta\vec k|)
        \right\}\;,
        \label{q91}
        \end{eqnarray}
where $J_0$ and $J_1$ are Bessel functions. This expression may be
further simplified by noting that $\vec k':=k \hat r$, i.e., $k=k'$.
Denoting the angle between $\vec k$ and $\vec k'$ by $\Theta$, one has:
        \[
        |\Delta\vec k|=2k\sin(\frac{\Theta}{2})=2k_x\;, ~~~~~~
        k_y^2-k_x^2=k^2\cos\Theta\;.\]
In view of  these relations, Eq.~(\ref{q91}), and making extensive use of the
properties of the Bessel functions \cite{gr} and the fact that
$F(r=0)=F(r=\infty)=1$, one finally arrives
at the following expression for the scattering amplitude:
	\begin{eqnarray}
        f^{(1)}(\vec k',\vec k) &=&
	\sqrt{2\pi k}e^{-3\pi i/4} \sin (\frac{\Theta}{2})
         \int_0^\infty dr (1-\frac{1}{F^2})
	 \left[
        -k\sin  (\frac{\Theta}{2})r \, J_0(2kr\sin \frac{\Theta}{2}) +
\right.\nonumber\\
	&& \left. ~~~~~~~~~~~  2(\lambda-\frac{1}{8\sin^2\frac{\Theta}{2}})\,
	 J_1(2kr\sin \frac{\Theta}{2})
        \right] \;.
        \label{q92}
        \end{eqnarray}
Here the term proportional to $\lambda$ signifies the contribution 
of the scalar curvature factor whereas the other terms reflect the effect of
the
kinetic energy term.

For the forward scattering ($\Theta=0$), this expression simplifies
to yield
        \begin{equation}
        f^{(1)}(\vec k,\vec k)=\sqrt{\frac{\pi}{8}}e^{-i3\pi/4} k^{3/2}\,
        \int_0^\infty dr'\left[ -r'(1-\frac{1}{F^2})\right],
        \label{q3.3}
        \end{equation}
As seen from Eq.~(\ref{q3.3}), the scalar curvature factor does not
contribute to the forward scattering.  It does however contribute to the
non-forward ($\Theta\neq 0$) scattering. For example consider the
back-scattering, where
	\[
	 f^{(1)}(\vec k'=-\vec k,\vec k)=\sqrt{2\pi k}e^{-3\pi i/4}
	\int_0^\infty dr
	(1-\frac{1}{F^2})\left[ -krJ_0(2kr)+2(\lambda-\frac{1}{8})J_1(2kr)\right] \;.
	\]

To see, in more detail, how the effects  due to the kinetic energy and the
scalar curvature terms compare, consider the Gaussian shape for the function
$f$, i.e., let
        \begin{equation}
        f(r)=\delta \, e^{-\mu r^2/2}\;,
        \label{q3.5}
        \end{equation}
where $\delta$ and $\mu$ are real parameters.
For convenience, let us also introduce the dimensionless parameter
$\eta:=\mu\,\delta^2$ and evaluate the integral in Eq.~(\ref{q92})
by first expanding the integrand in powers of $\eta$.  This involves integrals
of the form
	\begin{equation}
	\int_0^\infty r^ne^{-\alpha r^2}J_m(\nu r) dr\;, ~~~~~ (m=0,1, ~~ n\in{\ Z \hspace{-.08in}Z}^+, ~~
\alpha\in
	{\relax{\rm I\kern-.18em R}}^+)\:,
	\label{form}
	\end{equation}
which may be obtained using Ref.~\cite{gr}.

Carrying out the computations to the first  nonvanishing order in  $\eta$, one
has:
	\begin{eqnarray}
	 f^{(1)}(\vec k,\vec k)&=&\sqrt{\frac{2\pi}{k}}e^{-i3\pi/4}
	(\frac{k^2}{\mu})
	\left\{
	\left[
	\lambda-\frac{1}{4}-\frac{1}{8\sin^2\frac{\Theta}{2}}+
	\right.\right.\nonumber\\
	&&\left.\left.
	 \frac{k^2\sin^2\frac{\Theta}{2}}{4\mu}
	\right]
	\sin^2\frac{\Theta}{2}\,
	\exp(-\frac{k^2 \sin^2\frac{\Theta}{2}}{\mu})\:
	\eta + O(\eta^2)\right\}\;
	\label{q03.6}
	\end{eqnarray}
As seen from Eq.~(\ref{q03.6}), the scalar curvature and the kinetic energy
terms contributions to the scattering amplitude are comparable unless one
specializes to forward scattering.

\section{Effect of a Localized Magnetic Field} 
Consider the system of the previous section subject to a localized
cylindrically symmetric magnetic field $\vec B$. The latter may be 
defined by the vector potential $\vec A$ (connection one-form $A=A_rdr+A_\theta
d\theta +A_zdz$) with :
        \begin{eqnarray}
        A_r&:=&0\:=:\:A_z\nonumber\\
        A_\theta&:=& \frac{Br^2}{2}\, {\cal G}(r)\;,
        \label{q21}
        \end{eqnarray}
where ${\cal G}:[0,\infty)\to{\relax{\rm I\kern-.18em R}}$ is a smooth compactly supported function and
$B$ is a
constant parameter with the dimension of magnetic field. 

The Hamiltonian operator $\hat H_B$ for the constrained (two-dimensional) 
system subject to such a magnetic field is obtained by replacing $\hat p_i$ 
in Eq.~(\ref{q3}) by $\hat p_i+\frac{e}{c}\hat A_i$, where $e$ is the charge 
of the particle and $c$ is the speed of light. For the particular 
cylindrically symmetric system of interest, one finds:
        \begin{equation}
        \hat H_B=\hat H+\frac{e}{2mcr^2}(2\hat A_\theta \hat 
        p_\theta+ \frac{e}{c}\,\hat A_\theta^2)\;,
        \label{q22}
        \end{equation}
where $\hat H$ is the free Hamiltonian (\ref{q3}).

For convenience, let us denote the sum of the potential terms on 
the r.\ h.\ s.\ of (\ref{q22}) by $\Delta \hat V$. Expressing $\Delta\hat
V$ in the coordinate representation, where $\langle x|\hat p_\theta=-i\hbar
\frac{\partial}{\partial\theta}\langle x|$, one has:
        \begin{equation}
        \langle x|\Delta\hat V|k\rangle=\frac{e}{4\pi mcr^2}\left[
        2\hbar rA_\theta (k_y\cos \theta -k_x\sin \theta )+
        \frac{e}{c}\, A_\theta^2\right]\, e^{i\vec k.\vec x}\;.
        \label{q23}
        \end{equation}
It is this term that enters the expression for the scattering amplitude,
i.e., added to the terms due to $\hat H$. In view of Eqs.~(12), (\ref{q21}) and
(\ref{q22}), 
one needs to compute the integral:
        \begin{eqnarray} 
        \int d^2x'e^{-i\vec k'.\vec x}\langle x'|\Delta\hat V|k\rangle
        &=&\frac{e\hbar B}{4\pi mc}\int_0^\infty dr\int_0^{2\pi}d\theta
        \left[r^2{\cal G}(r)(k_y\cos\theta-k_x\sin\theta)
        e^{i(\vec k-\vec k').\vec x}\right]+\nonumber\\
        &&\frac{e^2B^2}{16\pi mc^2}\int_0^\infty dr\int_0^{2\pi}d\theta
        \left[ r^3{\cal G}^2(r)e^{i(\vec k-\vec k').\vec x}\right]\;.
        \label{q24}
        \end{eqnarray}
Choosing the $x$-axis in the $\Delta\vec k$ direction as in Eqs.~(19)
and (20), one can perform the angular integral. In terms of the angle
$\Theta$ between $\vec k$ and $\vec k'$, one has:
        \begin{eqnarray}
        \int d^2x'e^{-i\vec k'.\vec x}\langle x'|\Delta\hat V|k\rangle
        &=&\frac{e^2B^2}{8mc^2}\int_0^\infty dr
        r^3{\cal G}^2(r)\,J_0(2kr\sin\frac{\Theta}{2})+\nonumber\\
        &&\frac{ie\hbar Bk\cos(\Theta/2)}{2mc}\int_0^\infty dr
        r^2{\cal G}(r)\,J_1(2kr\sin\frac{\Theta}{2}). ~~~~
        \label{q201}
        \end{eqnarray}
The scattering amplitude is then obtained by adding (\ref{q201})
to the integral in Eq.~(12).

Now, consider a Gaussian shape for the function ${\cal G}$: 
        \begin{equation}
        {\cal G}(r)=e^{-r^2/2\sigma^2}.
        \label{q203}
        \end{equation}
In this case the integrals appearing in Eq.~(\ref{q201}) are again of the form
(\ref{form})
and easily evaluated. In view of Eq.~(12), one then has the following
expression for the contribution of magnetic field to the scattering amplitude:
	\begin{equation}
	\Delta f^{(1)}(\vec k',\vec k)=\sqrt{\frac{\pi}{2k}}e^{3\pi i/4}\left[
	\frac{e^2\Phi^2}{8\pi^2\hbar^2c^2}(1-k^2\sigma^2\sin^2\frac{\Theta}{2})
	+\frac{ie\Phi}{\pi\hbar
c}(k^2\sigma^2\sin\Theta)\right]e^{-\sigma^2k^2\sin^2\frac{
	\Theta}{2}}\;,
	\label{q202}
	\end{equation}
where
	\[
	\Phi:= \pi\sigma^2 B
	\]
is a characteristic magnetic flux\footnote{Note that the total magnetic flux is
zero as the topology of the space is ${\relax{\rm I\kern-.18em R}}^2$ and there are no singularities in
the fields.} and $i=\sqrt{-1}$.

Considering the Gaussian shape for both $f$ and ${\cal G}$, one obtains the
total scattering amplitude by adding the contributions of the geometry and
magnetic field:
	\begin{equation}
	f_B^{(1)}(\vec k',\vec k)=f^{(1)}(\vec k',\vec k)+\Delta f^{(1)}(\vec k',\vec
k)\;.
	\label{sum}
	\end{equation}
Here $f^{(1)}(\vec k',\vec k)$ is given by Eq.~(\ref{q03.6}).

Eq.~(\ref{sum}) may be used to give an order of magnitude estimate
of the size of the contribution of the scalar curvature factor. This
may be achieved by comparing the magnitudes of the two terms on the r.\ h.\ s.\
of this equation.  First note that for small magnetic
fluxes\footnote{Consideration of small magnetic fluxes is reasonable, because
here one tries to find a magnetic effect comparable with the effect of
geometry. The latter is an $\hbar^2$ order effect.} and for the case of
non-forward scattering which is of interest here, the term proportional to
$\Phi^2$ may be neglected.  Next for simplicity, choose $\mu =1/\sigma^2$ so
that the exponentially decaying factors in (\ref{q03.6}) and (\ref{q202}) are
the same. 
This  reduces the  comparison of the too effects to that of the following
terms:
	\begin{eqnarray}
	I)&
	2\left( \lambda-\frac{1}{4}-\frac{1}{8\sin^2\frac{\Theta}{2}}+
	 \frac{k^2\sin^2\frac{\Theta}{2}}{4\mu}
	\right)\eta
	\sin^2\frac{\Theta}{2}+O(\eta^2)&
	\label{term1}\\
	II)&\mbox{\large$\frac{e\Phi\sin\Theta}{\pi\hbar c}$}\:.&
	 \label{term2}
	\end{eqnarray}
Here used is made of the  choice $\mu =1/\sigma^2$. To simplify further,
consider the case of  electrons with $\eta\approx 10^{-1}$, $k \ll \sqrt{\mu}$
and $\Theta=\pi/2$.
Then a comparable magnetic effect has the following characteristic flux:
	\[
	\Phi
	\approx (\frac{\pi\hbar c\eta}{e})(\lambda-\frac{1}{2})
	\approx (\lambda-\frac{1}{2})\times
	10^{-8} ~~~~~~ ({\rm Gauss~Cm}^2) ;.
	\]
Thus an experimental setup capable of detection of the effect of the scalar
curvature factor
in scattering of free particles moving on a curved Gaussian shape surface
(\ref{q3.5}), must have a sensitivity to detect scattering of electrons due to
a localized magnetic field
 (\ref{q21}), (\ref{q203}) of characteristic flux   $\approx10^{-9}$ (Gauss
Cm$^2$) in the flat ${\relax{\rm I\kern-.18em R}}^2$.

\section{Discussion and Conclusion}
The physical consequences of the existence of a scalar curvature factor
in the scattering of free particles moving on a nonflat surface have been
analyzed. Although there has recently been some attempts to study
nonflat two-dimensional quantum systems, particularly in the context of
the quantum Hall effect \cite{nucl}, the curved spaces considered in the
literature are either spaces of constant scalar curvature \cite{b-s,aco}
or spaces with exotic topologies and geometries \cite{nucl}. On the other
hand in all these attempts, the possibility of the existence of
a scalar curvature factor in the Schr\"odinger equation has been ignored.

The particular system investigated in this article is physically more
interesting since there are indeed two-dimensional nonflat systems
with ${\relax{\rm I\kern-.18em R}}^2$ topology in nature. An example of this is the two-dimensional
electron system formed on the surface of liquid Helium 
\cite{isihara,edel}. Particularly remarkable is the Gaussian shape 
(\ref{q3.5}) of the surface of $ ^4\! H\! e$ in a dimple electron crystal in
the vicinity of the dimples \cite{isihara}.

To arrive at an experimental resolution of the factor ordering ambiguity,
i.e., experimental determination of the value of $\lambda$, a more
thorough investigation of the available (effectively) two-dimensional systems
is needed.

{\bf Note:} As indicated by the referee, in the comparison 
of the effect of the curvature  with the scattering effects of a cylindrically 
symmetric magnetic flux  in a flat two-dimensional system,  such a small
magnetic flux corresponds to a energy densities of the order $\sim 
10^{-17}$ ergs/Cm$^3$ whose effect  would likely be swamped by 
thermal effects in a $^4He$ system.  Here the analogy is used to give a 
very rough order of magnitude estimate for the maximum precision 
required for such an experiment. In practice one might not need 
such a precision.\footnote{For example in the above analysis the 
parameter $\eta$ is taken to be small only to ease the calculations.} In 
fact one may look at the collective effects such as those of a locally 
cylindrically symmetric curved surface, i.e.,  a surface curved at an array of 
points forming the vertices of a lattice. This is precisely the case in a dimple 
electron system. The main purpose of this comparison is to demonstrate 
that the corresponding effect is not several orders of magnitude beyond
the experimentally accessible values. This is usually the case where
the problems with the quantization on curved spaces are concerned.

\end{document}